\documentclass[a4paper,11pt]{article}
\pdfoutput=1
\usepackage{jheppub}
\usepackage[utf8]{inputenc}
\usepackage{epsfig,amsfonts,amsthm}
\usepackage{amsmath,amssymb}
\usepackage{float}
\usepackage{xcolor}

\newcommand{\be}{\begin{equation}}
\newcommand{\ee}{\end{equation}}
\newcommand{\bea}{\begin{eqnarray}}
\newcommand{\eea}{\end{eqnarray}}

\renewcommand{\Re}{\mathrm{Re }}
\renewcommand{\Im}{\mathrm{Im }}
\newcommand{\doublet}[2]{ \left( \begin{array}{c}#1 \\ #2 \end{array}\right) }

\newcommand{\triplet}[3]{ \left(\! \begin{array}{c}#1 \\ #2 \\ #3 \end{array}\!\right) }

\newcommand{\lr}[1]{ \langle #1 \rangle}

\newcommand{\Z}{\mathbb{Z}}
\newcommand{\mmatrix}[4]{ \left(\! \begin{array}{ccc}#1 & #2 \\ #3 & #4 \end{array}\!\right) }
\newcommand{\mmmatrix}[9]{ \left(\!\! \begin{array}{ccc}#1 & #2 & #3\\ #4 & #5 & #6\\ #7 & #8 & #9\\ \end{array}\!\!\right) }
\providecommand{\id}{{\boldsymbol{1}}}
\newcommand{\toCP}{\xrightarrow{CP}}
\newcommand{\hsm}{h_{\mbox{\tiny SM}}}
\newcommand{\msm}{m_{\mbox{\tiny SM}}^2}
\newcommand{\mmin}{m_{\mbox{\scriptsize min}}}
\newcommand{\mmax}{m_{\mbox{\scriptsize max}}}
\newcommand{\GeV}{\,\mbox{GeV}}

\newcommand{\lsim}{\mathrel{\rlap{\lower4pt\hbox{\hskip1pt$\sim$}}
		\raise1pt\hbox{$<$}}}         
\newcommand{\gsim}{\mathrel{\rlap{\lower4pt\hbox{\hskip1pt$\sim$}}
		\raise1pt\hbox{$>$}}}         
	
\definecolor{darkred}{rgb}{0.8,0.1,0.1}
\definecolor{paleyellow}{rgb}{1.0,1.0,0.7}

\title{Tridiagonal scalar mass matrix in the CP4 3HDM and its implications}

\author[a,1]{Bei Liu,}
\author[a,2]{Igor P. Ivanov,}
\author[b,c,3]{João Gonçalves}
\affiliation[a]{School of Physics and Astronomy, Sun Yat-sen University, 519082 Zhuhai, China}
\affiliation[b]{Department of Physics, Lund University, SE-223 62 Lund, Sweden}
\affiliation[c]{Departamento de F\'isica, Universidade de Aveiro and CIDMA, Campus de Santiago,\\ 3810-183 Aveiro, Portugal}
\emailAdd{liub98@mail2.sysu.edu.cn} 
\emailAdd{ivanov@mail.sysu.edu.cn} 
\emailAdd{jpedropino@ua.pt}

\abstract{
When parametrizing multi-Higgs potentials, it is desirable to express its coefficients
via observables. This is routinely done for the 2HDM, 
but this approach often fails in more elaborate models.
Here, we show that the scalar sector of the CP4 3HDM, an intriguing model based
on an order-4 $CP$ symmetry, can also be parametrized in an observable-driven manner.
The key feature that makes it work is the very special tridiagonal form of the $5\times 5$ 
neutral Higgs mass matrix.
We propose a set of input observables and present an algorithm to reconstruct 
the coefficients of the potential through linear relations.
Equipped with this procedure, we explore the scalar sector of the CP4 3HDM
beyond the limitations of previous studies. In particular, we identify a viable and testable regime 
in which all additional Higgses lie in the 300--600~GeV range.
This work offers a key ingredient for a future full phenomenological scan of this model.
}

\begin{document}
	\maketitle

\section{Introduction}

\subsection{Physics-driven parametrization of multi-Higgs sectors}

Multi-Higgs models \cite{Ivanov:2017dad} represent a phenomenologically rich framework 
for constructions beyond the Standard Model (BSM).
Such models often possess a Standard-Model-like (SM-like) scalar $\hsm$, identified with the 125 GeV particle 
found at the LHC \cite{ATLAS:2012yve,CMS:2012qbp},
together with several additional Higgs bosons, which are yet to be discovered. 
When building such models, one tries to make sure that the couplings of $\hsm$ deviate from the SM predictions 
in a controlled way and that the additional scalars are either heavy enough or possess suppressed couplings to the SM fields.

Naively, one can start with a lagrangian, which contains numerous coefficients in its scalar potential and Yukawa sector, 
and compute the masses and other properties of all the physical scalars in terms of these coefficients.
This straightforward procedure of exploring the model via a random scan in the parameter space
is often computer-time expensive and offers little physical insight, as each new random set of parameters
may lead to a completely different phenomenological scenario.

A much more insightful procedure is to take some of the physical observables 
as input and reconstruct the coefficients of the potential or Yukawa interactions in terms of these observables.
One then proceeds with making further predictions, which are expressed in terms of measurable quantities.
In this way, exploration of the model can be guided by desired phenomenological features.
We call this procedure ``inversion'' because it inverts the traditional sequence ``coefficients $\to$ observables''.

In simple multi-Higgs models, such as the very popular two-Higgs-doublet model (2HDM) 
with a softly broken $\Z_2$ symmetry \cite{Branco:2011iw}, inversion can be carried out in a straightforward way.
The coefficients of the potential $m_{11}^2$, $m_{22}^2$, and five $\lambda$'s can be
traded for the overall vacuum expectation value (vev) $v$, the mixing angles $\alpha$ and $\beta$, 
and the masses of all the physical scalars.
This is why exploration of this model is so convenient: one can plot the other observables
against the physical parameters such as $\tan\beta$, $\cos(\beta-\alpha)$, or $m_H$.
Even in the general 2HDM, with its 11 relevant free parameters in its scalar sector,
this physics-driven parametrization of the potential can be carried out, making use of 
certain triple and quartic scalar vertices \cite{Grzadkowski:2013rza,Ginzburg:2015cxa,Ferreira:2020ana}.

Beyond two Higgs doublets, and especially in models with additional global symmetries,
implementation of this idea runs into technical difficulties.
First, with many new scalar fields, possible observables such as scalar masses 
and mixing angles quickly outnumber free parameters.
Moreover, the main difficulty is not to choose a subset of observables to serve as input,
but rather to make sure that these observables can indeed be chosen independently, and within which ranges.
This task is aggravated by the fact that scalar masses and mixing angles
are expressed via the coefficients of the lagrangian in a complicated non-linear way,
which is not guaranteed to be analytically invertible.

To give a concrete example, consider three-Higgs-doublet models (3HDM), another popular and rich
example of non-minimal Higgs sector \cite{Weinberg:1976hu,Ivanov:2017dad}.
For a normal, neutral minimum, it contains five neutral scalars and two pairs of charged Higgs bosons.
Within the neutral Higgs sector alone, we have five Higgs boson masses and ten rotation angles 
which diagonalize the neutral scalar mass matrix.
If the scalar sector contains enough freedom, one could use the ten rotation
angles as input parameters and use them to express the coefficients of the potential, 
as it was recently done, for example, in the 3HDM with a softly broken $\Z_2\times \Z_2$ symmetry \cite{Boto:2024jgj}.
However, many 3HDM with global symmetries often have fewer free parameters,
in which case the scalar masses and mixing angles exhibit non-trivial correlations, 
which are not easy to identify.
Even if one focuses on the Higgs spectrum alone, there is no guarantee that 
{\em any} choice of scalar masses can be accommodated within the model one studies.
This is especially true for symmetry-based models without the decoupling limit \cite{Faro:2020qyp,Carrolo:2021euy}.

In such situations, one is often tempted to return to the traditional procedure:
perform a random scan of the parameter space, compute the observables numerically for each point, 
and simply select only those points which have those observables within the desired ranges.
However, if the number of free parameters is large, this poor man's approach is computer-time consuming and 
extremely inefficient. A better procedure is definitely needed.

\subsection{The goals of the present work}

In this work, we address this task within the CP4 3HDM, 
the intriguing 3HDM based on the exact $CP$ symmetry of order 4, 
which was first described in \cite{Ivanov:2015mwl}. 
The scalar sector of this model is rather elaborate, with 13 free parameters.
The first phenomenological study of this model
undertaken in \cite{Ferreira:2017tvy} was based on a traditional random scan in the parameter space
for scalar and Yukawa sectors. Even with the additional assumption of the exact scalar alignment,
the scan was rather inefficient. Moreover, virtually all the parameter space points
identified in \cite{Ferreira:2017tvy} were later shown to conflict the LHC Run 2 data \cite{Ivanov:2021pnr}.

However, the recent work \cite{Zhao:2023hws} succeeded in implementing the inversion procedure
in the Yukawa sector of the CP4 3HDM.
Even without details on the scalar sector, this procedure spectacularly demonstrated 
that only two out of eight possible CP4-invariant Yukawa scenarios have the chance 
of producing viable benchmark models.
These Yukawa sectors must now be coupled with the CP4 3HDM scalar sector 
which would use as much physics input as possible, and then a full phenomenological scan 
of the model should be undertaken.

Thus, the main motivation for this work is to be able to parametrize the scalar sector of the CP4 3HDM 
in a physics-driven way, to be able to perform a full phenomenological analysis in a follow-up paper.

We found that a physics-based parametrization of the scalar sector 
can indeed be constructed for the elaborate scalar sector of the CP4 3HDM.
The observables used for physics input are the vevs and the properties of the SM-like Higgs $\hsm$.
This key observation is that the neutral $5\times 5$ mass matrix
is {\em tridiagonal}: it only contains non-zero elements on the main diagonal 
and on the parallel diagonals one step above and below. 
It is this non-generic feature of a non-block-diagonal matrix that allows us to 
construct a {\em linear} inversion procedure without the need to solve non-linear equations.
Demonstrating this strategy is the first main goal of this paper.

Curiously, we found that even this physics-driven parametrization has its own pitfalls. 
We will show that our initial implementation of this strategy missed a phenomenologically 
important regime of this model, and we will explain how we discovered and corrected the flaw.
Demonstration of these subtle issues is the second goal of this work.
We believe that the lessons learned with this example may prove useful 
for other multi-Higgs sectors.

The third major goal of the paper is to actually explore the scalar sector of the CP4 3HDM
beyond the assumptions and restrictions of Ref.~\cite{Ferreira:2017tvy}.
In particular, we will show how the minimal $\mmin$ and the maximal $\mmax$ masses 
of the extra four neutral scalars depend on the input parameters.
In a future publication, these results, together with the viable Yukawa sectors 
of this model found in \cite{Zhao:2023hws},
will be used in a detailed phenomenological scan of the model.

The structure of the paper is the following.
In Section~\ref{section-CP4-scalar} we give explicit analytical expressions 
for the scalar mass matrices of the CP4 3HDM. These calculations
partly repeat the results of Ref.~\cite{Ferreira:2017tvy} but we now stress
the all-important tridiagonal form of the mass matrix not noticed before.
Next, in Section~\ref{section-inversion} we explain how this tridiagonal form
makes possible the linear inversion of the scalar potential.
We describe here the choice of the input parameters and outline the procedure.
Section~\ref{section-numerical} presents the numerical exploration of the problem.
We first show a general scan, identify a subtle problem, resolve it,
and show a new, ``focused'' scan which allows us to reach reasonably high masses
for all additional Higgses.
We conclude the paper with a brief discussion and an outlook.
Technical details are delegated to the appendices.


\section{The scalar sector of CP4 3HDM}\label{section-CP4-scalar}

\subsection{The CP4 3HDM story so far}

The 3HDM activity is driven by the realization that by merely extending the concept of three generations
from the fermion to the scalar sector and equipping the model with (approximate) global symmetries
one can generate many phenomenological and cosmological scenarios, which are not possible with 
more restricted scalar sector such as the 2HDM \cite{Branco:2011iw} or singlet extensions.
From the model building perspective, the symmetry options available within the 2HDM \cite{Ivanov:2006yq,Nishi:2006tg,Ivanov:2007de,Maniatis:2007de,Ferreira:2009wh,Cogollo:2016dsd,Alves:2018kjr}
are nowhere near the wealth of symmetry-based situations that can be implemented in the 3HDM,
in the scalar sector alone \cite{Ivanov:2011ae,Ivanov:2012ry,Ivanov:2012fp,Darvishi:2019dbh} 
or together with the Yukawa interactions \cite{Ferreira:2010ir, Ivanov:2013bka, Bree:2023ojl}.

A vivid illustration of this trend is given by so-called generalized, or general, $CP$ symmetries (GCPs).
It is known since long ago that the $CP$ transformation is not uniquely defined in quantum field theory \cite{feinberg-weinberg,Lee:1966ik,Branco:1999fs,weinberg-vol1},
and this freedom of definition increases further in models with several complex fields with equal quantum numbers.
For example, a GCP acting on complex scalar fields $\phi_i$, $i = 1, \dots, N$,
can be written as \cite{Ecker:1987qp,Grimus:1995zi}
\begin{equation}
	\phi_i({\bf r}, t) \toCP {\cal CP}\,\phi_i({\bf r}, t)\, ({\cal CP})^{-1} = X_{ij}\phi_j^*(-{\bf r}, t), \quad X_{ij} \in U(N)\,.
	\label{GCP}
\end{equation}
The conventional definition of $CP$ with $X_{ij} = \delta_{ij}$ is only one of many possible
choices and is, in fact, basis-dependent.
If a model does not respect the conventional $CP$ but is invariant under a GCP with a suitable matrix $X$,
then the model is explicitly $CP$-conserving \cite{Branco:1999fs}.

The presence of the matrix $X$ has consequences.
Applying the GCP twice leads to a family transformation, $\phi_i \mapsto (XX^*)_{ij}\phi_j$,
which may be non-trivial.
One may find that only when the GCP transformation is applied $k$ times that one arrives at the identity transformation;
thus, a $CP$ symmetry can be of order $k$.
The conventional $CP$ is of order 2; the next option is a $CP$ symmetry of order 4, denoted CP4.
Since the order of a transformation is basis-invariant, a model based on a CP4
represents a physically distinct $CP$-invariant model which cannot be achieved with the conventional $CP$.

Within the 2HDM, imposing CP4 on the scalar sector always leads to the usual $CP$ \cite{Ferreira:2009wh}.
One can extend CP4 to the Yukawa sector \cite{Maniatis:2009vp}, but the resulting model is severely constrained 
and leads to pathologies in the quark sector.
Within the 3HDM context, this unconventional $CP$ symmetry leads to a remarkable and apparently viable model.
Dubbed CP4 3HDM, it was explicitly constructed in \cite{Ivanov:2015mwl} 
and was found to possess remarkable features which sometimes defy intuition \cite{Ivanov:2015mwl,Aranda:2016qmp,Haber:2018iwr}.
If CP4 remains unbroken at the minimum of the Higgs potential, it can protect the scalar dark matter candidates against decay
\cite{Ivanov:2018srm} and may be used to generate radiative neutrino masses \cite{Ivanov:2017bdx}.
CP4 symmetry can also be extended to the Yukawa sector leading to very particular patterns of the Yukawa matrices 
\cite{Ferreira:2017tvy,Zhao:2023hws}.
The CP4 transformation, by construction, mixes generations; therefore,
in order to avoid mass degenerate quarks, the explicit CP4 symmetry of the model must be spontaneously broken.
Then, the Yukawa sector contains enough free parameters to accommodate
the experimentally measured quark masses, mixing, and $CP$ violation.
In fact, it was recently demonstrated in \cite{Zhao:2023hws} that the Yukawa sector of the CP4 3HDM
can be inverted: namely, one can use the experimentally known quark masses 
and the Cabibbo-Kobayashi-Maskawa (CKM) matrix as input and, for any vacuum expectation values of the Higgs doublets,
reconstruct the original Yukawa coupling matrices in the lagrangian. 
This inversion procedure significantly sped up exploration of the model 
as compared with the old scan \cite{Ferreira:2017tvy}.

A generic multi-Higgs model leads to tree-level Higgs-mediated flavor changing neutral couplings (FCNC),
which are severely constrained by the neutral meson oscillation parameters. 
This fact prompted many model builders to eliminate the tree-level FCNCs altogether by 
implementing the natural flavor conservation principle~\cite{Glashow:1976nt,Paschos:1976ay,Peccei:1977hh}.
This assumption is not compulsory; certain amount of tree-level FCNCs
can be tolerated when sufficiently suppressed by structural features of the models,
see \cite{Sher:2022aaa} for a review of various options.

Within the CP4 3HDM, due to its links between the scalar and Yukawa sectors, natural flavor conservation
cannot be imposed, and the tree-level Higgs-mediated FCNCs seem unavoidable.
The recent study \cite{Zhao:2023hws} explored the issue in detail and arrived at a remarkable conclusion:
out of eight possible CP4-invariant quark sectors, only two have a chance to pass all the constraints
and lead to viable models.
Remarkably, this strong conclusion was obtained from the Yukawa sector alone, 
without performing scans in the scalar potential parameter space:
FCNCs in the other scenarios were so strong that the predictions 
would severely conflict the data for any CP4 3HDM scalar sector.

\subsection{The potential}

The three Higgs doublets of the 3HDM, $\phi_i$ with $i = 1,2,3$, have identical quantum numbers.
As defined in Eq.~\eqref{CP4-def}, the transformation CP4 acts on Higgs doublets by complex conjugation 
accompanied by a unitary matrix $X$ which must satisfy $(XX^*)^2 = 1$.
As known since long ago \cite{Ecker:1987qp,Grimus:1995zi}, by performing a basis change, 
one can bring the matrix $X$ to a simple block-diagonal form.
As suggested in \cite{Ivanov:2015mwl,Ferreira:2017tvy}, we use the following form of the CP4:
\begin{equation}
	\phi_i \toCP X_{ij} \phi_j^*\,,\quad
	X =  \left(\begin{array}{ccc}
		1 & 0 & 0 \\
		0 & 0 & i  \\
		0 & -i & 0
	\end{array}\right)\,.
	\label{CP4-def}
\end{equation}
Note that applying this transformation twice leads to a non-trivial transformation in the space of doublets:
$\phi_{1} \mapsto \phi_{1}$, $\phi_{2,3} \mapsto -\phi_{2,3}$.

The 3HDM scalar sector invariant under CP4 was constructed in \cite{Ivanov:2015mwl}. 
As suggested in \cite{Ferreira:2017tvy}, one can use the residual basis change freedom 
that leaves the matrix $X$ invariant to eliminate 
a few redundant free parameters, resulting in the following form of the potential:
\begin{eqnarray}
	V_0 &=& - m_{11}^2 (\phi_1^\dagger \phi_1) - m_{22}^2 (\phi_2^\dagger \phi_2 + \phi_3^\dagger \phi_3) \nonumber\\
	&&+ \lambda_1 (\phi_1^\dagger \phi_1)^2 + \lambda_2 \left[(\phi_2^\dagger \phi_2)^2 + (\phi_3^\dagger \phi_3)^2\right]
	+ \lambda_{34} (\phi_1^\dagger \phi_1) (\phi_2^\dagger \phi_2 + \phi_3^\dagger \phi_3) \nonumber\\ 
	&&- \lambda_4 \left[(\phi_1^\dagger \phi_1) (\phi_2^\dagger \phi_2) - (\phi_1^\dagger \phi_2)(\phi_2^\dagger \phi_1)
	+ (\phi_1^\dagger \phi_1) (\phi_3^\dagger \phi_3) - (\phi_1^\dagger \phi_3)(\phi_3^\dagger \phi_1)\right]\nonumber\\
	&&+
	\lambda'_{34} (\phi_2^\dagger \phi_2) (\phi_3^\dagger \phi_3)
	- \lambda'_4 \left[(\phi_2^\dagger \phi_2) (\phi_3^\dagger \phi_3) - (\phi_2^\dagger \phi_3)(\phi_3^\dagger \phi_2)\right]\,,
	\label{V0}
\end{eqnarray}
with all parameters real, and
\begin{equation}
	V_1 = \lambda_5 (\phi_3^\dagger\phi_1)(\phi_2^\dagger\phi_1) +
	\lambda_8(\phi_2^\dagger \phi_3)^2 + \lambda_9(\phi_2^\dagger\phi_3)(\phi_2^\dagger\phi_2-\phi_3^\dagger\phi_3) + h.c.
	\label{V1a}
\end{equation}
with real $\lambda_5$ and complex $\lambda_8$, $\lambda_9$.
This parametrization contains 13 real free parameters.
Compared to the parametrization used in \cite{Ferreira:2017tvy},
we slightly changed the parametrization:
instead of $\lambda_{3}$ and $\lambda_{3}'$, we now use $\lambda_{34} = \lambda_3+\lambda_4$,
and $\lambda_{34}' = \lambda_3'+\lambda_4'$, while keeping $\lambda_4$ and $\lambda_4'$.
In this way, $\lambda_4$ and $\lambda_4'$ disappear from the neutral Higgs mass matrices 
and can be fixed by the choice of the charged Higgs masses.
As proved in \cite{Ivanov:2015mwl}, this potential has no other symmetries 
apart from the transformations that can be obtained via successive applications of CP4.

The hallmark feature of the CP4 3HDM is that, although the model is explicitly $CP$-conserving,
there is no basis in which all the coefficients of the potential are real.
The terms in Eq.~\eqref{V1a} contain two complex parameters, $\lambda_8$ and $\lambda_9$.
By performing rephasing transformations, one could render either $\lambda_8$ or $\lambda_9$ real 
but not both of them simultaneously.

\subsection{The minimum conditions and mass matrices}

The minimum of the Higgs potential can either preserve or break CP4. 
The former option was explored in \cite{Ivanov:2018srm} to build the CP4-stabilized dark matter model.
However, if one extends the CP4 symmetry to the Yukawa sector, the unbroken CP4 leads to 
mass-degenerate quarks \cite{Aranda:2016qmp,Ferreira:2017tvy} which is clearly in conflict with experiment.
Thus, we must consider the situation when CP4 is spontaneously broken.
It was shown in \cite{Ferreira:2017tvy} that, with a suitable basis change, 
one can always make the vevs real and parametrize them using two angles $\beta$ and $\psi$
(below, $c_x$ and $s_x$ stand for cosine and sine of any angle $x$):
\begin{equation}
	\lr{\phi_i^0} = \frac{1}{\sqrt{2}} (v_1,\, v_2,\, v_3) \equiv
	\frac{v}{\sqrt{2}} (c_\beta,\, s_\beta c_\psi,\, s_\beta s_\psi) \,.
	\label{vevs1}
\end{equation}
The angle $\beta$ plays a role similar to that in the $Z_2$-symmetric 2HDM: 
it shows the mixing between the Higgs doublets 
which transform trivially and non-trivially under the scalar sector symmetry.
The angle $\psi$ is a new quantity which describes the rotation inside the $(\phi_2,\phi_3)$ block.

Note that spontaneous breaking of CP4 leads to existence of four distinct minima
which are related by the CP4 transformation  but which are physically equivalent.
As a result, without loss of generality, one can choose $0 \le \psi \le \pi$ so that $\sin\psi \ge 0$.
However the sign of $\cos\psi$ is not yet fixed and, as well see later, is correlated with
the sign of $\lambda_5$.

We can expand the three doublets around the minimum as
\begin{equation}
	\phi_1 = {1\over\sqrt{2}}\doublet{\sqrt{2}h_1^+}{vc_\beta + h_1 + i a_1},\
	\phi_2 = {1\over\sqrt{2}}\doublet{\sqrt{2}h_2^+}{vs_\beta c_\psi + h_2 + i a_2},\
	\phi_3 = {1\over\sqrt{2}}\doublet{\sqrt{2}h_3^+}{vs_\beta s_\psi + h_3 + i a_3},\label{expansion}
\end{equation}
and derive the extremum conditions (also known as tadpole conditions) by requiring that the linear terms be absent.
Due to the real-valued vevs, the conditions on $a_i$ are satisfied when the imaginary parts 
of $\lambda_8$ and $\lambda_9$ obey the simple relation
	$s_{2\psi} \Im (\lambda_8) + c_{2\psi}\Im (\lambda_9) = 0$.
It allows us to parametrize the two imaginary parts as
\begin{equation}
	\Im(\lambda_8) = c_{2\psi} \lambda_{89}\,, \quad \Im(\lambda_9) = -s_{2\psi} \lambda_{89}\,,
	\quad
	\lambda_{89} \equiv \sqrt{(\Im \lambda_8)^2 + (\Im\lambda_9)^2}\,.
	\label{lam89}
\end{equation}
The tadpole equations for $h_i$ lead to 
\begin{eqnarray}
\frac{m_{11}^2}{v^2} &=& \lambda_1 c_\beta^2 + {1 \over 2} \lambda_{34} s_\beta^2 
+ {1 \over 2} \lambda_5 s_\beta^2\, s_{2\psi}\,,
\label{m11}\\
\frac{m_{22}^2}{v^2} &=& \lambda_2 s_\beta^2 + {1 \over 2}\lambda_{34} c_\beta^2
+ {1\over 2}  \Re(\lambda_9) s_\beta^2\, t_{2\psi}\,,
\label{m22}\\
0 &=&\lambda_5 c_\beta^2 c_{2\psi} + \Lambda s_\beta^2\, s_{2\psi} c_{2\psi} 
+ \Re (\lambda_9)s_\beta^2\, c_{4\psi}\,,
\label{t23}
\end{eqnarray}
where we used the following shorthand notation:
\begin{equation}
	\Lambda \equiv {\lambda'_{34}\over 2} + \Re (\lambda_8) - \lambda_2\,.
	\label{Lambda}
\end{equation}
The Higgs mass matrices, ${\cal M}_{ch}$ for the charged sector and ${\cal M}_n$ for the neutral sector, 
can be calculated explicitly in this basis, 
\cite{Ferreira:2017tvy};
for completeness, we present them in Appendix~\ref{appendix-mass-matrices}.

\subsection{The Higgs basis}

Description of the scalar sector becomes more transparent 
if we pass from the original doublets $\phi_i$ to a Higgs basis $\Phi_i$, where only one doublet gets a non-zero vev:
$\lr{\Phi_1^0} = v/\sqrt{2}$, $\lr{\Phi_2} = \lr{\Phi_3} = 0$.
In particular, the Higgs basis plays an important role in the derivation and the analysis
of the FCNC matrices \cite{Zhao:2023hws}.
Since the choice of the Higgs basis is not unique in the 3HDM, we use the Higgs basis defined 
by the following transformation of the doublets $\phi_i$ to $\Phi_i$:
\begin{equation}
	\Phi_i = {\cal P}_{ij} \phi_j\,, \qquad
	\triplet{\Phi_1}{\Phi_2}{\Phi_3} =
	\mmmatrix{c_\beta}{s_\beta c_\psi}{s_\beta s_\psi}{-s_\beta}{c_\beta c_\psi}{c_\beta s_\psi}{0}{-s_\psi}{c_\psi}
	\triplet{\phi_1}{\phi_2}{\phi_3}\,, 
	\label{Phiphi}
\end{equation}
Here, we follow the notation of \cite{Zhao:2023hws}; it differs from the convention used in the earlier paper \cite{Ferreira:2017tvy} 
by the exchange of the second and third doublets $\Phi_{2,3}$.

As we will see below, the particular choice of the Higgs basis used in Eq.~\eqref{Phiphi} does not diagonalize the charged Higgs mass matrix.
Diagonalization can be achieved by performing an additional rotation between $\Phi_2$ and $\Phi_3$, 
which would bring us to the so-called charged Higgs basis \cite{Bento:2017eti}.
Although this choice would simplify the form of the charged Higgs sector, it would spoil a very helpful feature 
in the neutral Higgs sector --- the tridiagonal mass matrix structure to be shown below --- which proves instrumental to our analysis.
This is why we stand with the choice of the basis presented in Eq.~\eqref{Phiphi}.
In this Higgs basis, we expand the doublets with the following notation:
\begin{equation}
	\Phi_1 = {1\over\sqrt{2}}\doublet{\sqrt{2}G_1^+}{v + \rho_1 + i G^0},\
	\Phi_2 = {1\over\sqrt{2}}\doublet{\sqrt{2}w_2^+}{\rho_2 + i \eta_2},\
	\Phi_3 = {1\over\sqrt{2}}\doublet{\sqrt{2}w_3^+}{\rho_3 + i \eta_3}.\label{expansion-Higgs-basis}
\end{equation}
The transformation of the charged degrees of freedom is given by the same matrix ${\cal P}_{ij}$.
As for the neutral fields, it is convenient to group them in the six dimensional vector of real fields
\begin{equation}
	\varphi_a = (h_1, h_2, h_3, a_1, a_2, a_3) \quad\mbox{and}\quad \Phi_a = (G^0, \rho_1, \rho_2, \rho_3, \eta_3, \eta_2)\,.
	\label{Higgs-r}
\end{equation}
The ordering choice for the last two components of $\Phi_a$ ($\eta_3, \eta_2$, not $\eta_2, \eta_3$)
is driven by convenience and will be explained shortly.
The explicit relation between $\varphi_a$ and $\Phi_a$ is 
\begin{equation}
	\Phi_a = P_{ab}\,\varphi_b\,, \quad
	P_{ab} = \left(\begin{array}{cccccc}
		0 & 0 & 0 & c_\beta & s_\beta c_\psi  & s_\beta s_\psi  \\[1mm]
		c_\beta & s_\beta c_\psi  & s_\beta s_\psi  & 0 & 0 & 0 \\[1mm]
		-s_\beta & c_\beta c_\psi  & c_\beta s_\psi & 0 & 0 & 0 \\[1mm]
		0 & - s_\psi  & c_\psi  & 0 & 0 & 0 \\[1mm]
		0 & 0 & 0 & 0 & - s_\psi  & c_\psi \\[1mm]
		0 & 0 & 0 & -s_\beta & c_\beta c_\psi  & c_\beta s_\psi  
	\end{array}\right)\,.\label{rotation-matrix-P}
\end{equation}
In the Higgs basis, the charged scalar mass terms become
$h_i^- ({\cal M}_{ch})_{ij} h_j^+ = w_i^- ({\cal M}^{H}_{ch})_{ij} w_j^+ $, 
where $w_i^+ = (G^+, w_2^+, w_3^+)$,
with the following mass matrix:
\begin{equation}
	{\cal M}^{H}_{ch} = {v^2\over 2}\mmmatrix{0}{0}{0}
	{0}{-\lambda_4 -\lambda_5 s_{2\psi}}%
	{-\lambda_5 c_\beta c_{2\psi}}%
	{0}{-\lambda_5 c_\beta c_{2\psi}}%
	{-\lambda_4 c_\beta^2 +\lambda_5 c_\beta^2 s_{2\psi} + \tilde\Lambda s_\beta^2}\,,
	\label{mass-matrix-charged-Higgs}
\end{equation}
where $\tilde\Lambda$ is 
\begin{eqnarray}
	\tilde\Lambda \equiv \lambda'_{34} - \lambda'_4 - 2\lambda_2 - 2 \Re(\lambda_9)\, t_{2\psi}\,.
	\label{tildeLam}
\end{eqnarray}
The neutral mass terms are
	$\varphi_a^T ({\cal M}_n)_{ab} \varphi_b = \Phi^T_a {\cal M}^{H}_{ab} \Phi_b$,
so that the neutral mass matrix in the Higgs basis is ${\cal M}^{H}= P {\cal M}_n P^T$.
The direct calculation shows that it takes, rather unexpected, 
the following tridiagonal form:
\begin{equation}
	{\cal M}^{H} = \mmatrix{0}{\vec 0}{\vec 0}{{\cal \tilde M}}\,,
	\quad 	
	{\cal \tilde M} = \left(\begin{array}{ccccc}
		a_{11} & a_{12} & 0 & 0 & 0  \\[1mm]
		a_{12} & a_{22} & a_{23} & 0 & 0  \\[1mm]
		0 & a_{23} & a_{33} & a_{34} & 0  \\[1mm]
		0 & 0 & a_{34} & a_{44} & a_{23}  \\[1mm]
		0 & 0 & 0 & a_{23} & a_{55} 
	\end{array}\right)\,.\label{M-Higgs-basis}
\end{equation}
As expected, the first row and column in ${\cal M}^{H}$ are filled with zeros 
as they correspond to the would-be Goldstone boson $G^0$,
and the remaining $5\times 5$ matrix ${\cal \tilde M}$ acts in the space 
$(\rho_1, \rho_2, \rho_3, \eta_3,\eta_2)$.
Our choice of the ordering $\eta_3,\eta_2$ was dictated by the desire to present this matrix in the tridiagonal form.

Let us now list the elements of ${\cal \tilde M}$ and discuss their roles.
The first row was already given in \cite{Ferreira:2017tvy}: 
\begin{equation}
	a_{11} = 2c_\beta^2 m_{11}^2 + 2s_\beta^2 m_{22}^2 \,,\quad
	a_{12} = -\sin{2\beta}\, (m_{11}^2 - m_{22}^2)\,.
	\label{aa-11-12}
\end{equation}
We conclude that if $m_{11}^2 = m_{22}^2$, we get exact scalar alignment without decoupling.
The SM-like Higgs $H_{125}$ would be identical to $\rho_1$, and its mass $m_{H_{125}}^2 = 2m_{11}^2$.
This is the only way the exact scalar alignment can be imposed on this model, 
as $\sin2\beta = 0$ would lead to unphysical quark sector when CP4 is extended to the fermion sector.

The remaining elements in the $\rho_i$ block are 
\begin{eqnarray}
	a_{22} &=& 2s_\beta^2 m_{11}^2 + 2c_\beta^2 m_{22}^2 - v^2(\lambda_{34} + \lambda_5 \sin2\psi)\,,\nonumber\\
	a_{23} &=& - \lambda_5 v^2 c_\beta \cos2\psi\,,\nonumber\\
	a_{33} &=& \lambda_5 v^2 c_\beta^2 \sin2\psi + (\Lambda - 2\Re\lambda_9\,  t_{2\psi}) v^2s_\beta^2\,.
	\label{aa-rho}
\end{eqnarray}
The $(\eta_3,\eta_2)$ block contains
\begin{equation}
	a_{44} = \lambda_5 v^2 c_\beta^2 \sin2\psi + \Lambda' v^2 s_\beta^2\,,\quad
	a_{45} = -\lambda_5 v^2 c_\beta \cos2\psi\,,\quad
	a_{55} = -\lambda_5 v^2 \sin2\psi\,,
	\label{aa-eta}
\end{equation}
where $\Lambda'$ is 
\begin{equation}
	\Lambda' \equiv \Lambda - 2\Re (\lambda_8) - \Re(\lambda_9)\, t_{2\psi} = 
	{\lambda'_{34}\over 2} - \Re (\lambda_8) - \lambda_2 - \Re(\lambda_9)\, t_{2\psi}\,.\label{LamLam}
\end{equation}
Note that $a_{45} = a_{23}$ and $a_{55}/a_{45} = t_{2\psi}/c_\beta$.
Also, the expression for $a_{55}$ makes it clear that the sign of $\cos\psi$ is correlated 
with the sign of $\lambda_5$: if we choose $0 \le \psi \le \pi/2$ so that $\cos\psi > 0$,
then we must impose $\lambda_5 < 0$. 

Finally, mixing between the real and imaginary parts of the neutral fields arises
through a single term:
\begin{equation}
	a_{34} = -\lambda_{89} s_\beta^2 v^2\,.
	\label{aa-34}
\end{equation}
Had we started with a model with real $\lambda_8$ and $\lambda_9$,
we would find that the model possessed not only CP4 but also the usual $CP$. 
Then the model would also possess the accidental $D_4$ Higgs family symmetry group, 
which would forbid spontaneous $CP$ violation in the scalar potential of the 3HDM.

\section{Physics-driven parametrization of the CP4 3HDM}\label{section-inversion}

\subsection{The problem}

The CP4 3HDM potential \eqref{V0}, \eqref{V1a}, contains in total 13 real free parameters:
$m_{11}^2$, $m_{22}^2$, and various $\lambda$'s.
This number is reduced to 12 if we fix the real vev basis.
In principle, the vevs of all doublets and the physical Higgs boson masses and couplings 
come from these parameters. 
However, since one must deal with the $5\times 5$ neutral Higgs mass matrix diagonalization,
these expressions cannot be written in analytical form.
One can perform a random scan in the space of $\lambda$'s, as was indeed done 
in the first phenomenological study of the CP4 3HDM \cite{Ferreira:2017tvy}, 
but the procedure was phenomenologically inefficient:
each random point resulted in vastly different phenomenological regimes.

As described in the introduction, we now aim to construct the physics-driven 
parametrization of the scalar sector of the CP4 3HDM.
That is, we choose a set of quantities which are linked with observables
and algorithmically express the coefficients of the potential in terms of these quantities.

Before we present our strategy, we would like to stress that this task is non-trivial 
due to a potential non-linearity.
Suppose we consider the $5\times 5$ neutral mass matrix and choose the vevs and the five 
masses as the main input parameters.
Does it imply that we are free to select any values of these masses?
Definitely not. First of all, the model does not have the decoupling limit,
and second, the relations between the coefficients and the Higgs masses are highly non-linear.
Therefore, the possible values of masses and vevs available in the CP4 3HDM
fill a particular region in the parameter space of masses and vevs.
The shape and size of this region is {\em \`{a} priori} unknown.
As a result, we do not know how to choose a suitable set of masses and how to express 
the coefficients in terms of masses.

What we should strive for is to establish a {\em linear} algorithm. 
That is, we need a set of observables or quantities which have a clear physical meaning, 
which will allow us to express the elements of the mass matrix and the coefficients of the potential 
via linear relations.
In this way, the strategy can be always implemented without solving non-linear equations. 

Of course, once the algorithm is implemented, it may lead to a set of coefficients 
which eventually violate one or several constraints on the scalar potential.
This is unavoidable, and the algorithm must be complemented with such posterior checks.
The key point is that the algorithm manages to uniquely and straightforwardly build the potential
starting from the chosen set of observables.

\subsection{A generic inversion algorithm}

In \cite{Zhao:2023hws}, the inversion procedure was successfully constructed
for the Yukawa sector of the CP4 3HDM. That procedure used as input the three quantities coming from the scalar vevs: 
$v = 246$~GeV and the two angles, $\beta$ and $\psi$. 
They must, of course, be retained as input parameters 
in the physically motivated inversion algorithm we are now constructing.

Next, the structure of the charged Higgs mass matrix Eq.~\eqref{mass-matrix-charged-Higgs}
contains $\lambda_4$ and $\lambda_4'$ on the diagonal. It allows us to choose the two charged Higgs masses
as input parameters and to express $\lambda_4$ and $\lambda_4'$ through them and the other parameters.

Out of the five neutral scalar bosons, we single out the SM-like Higgs $\hsm$ and use its properties as input parameters.
In contrast to the earlier papers \cite{Ferreira:2017tvy,Zhao:2023hws}, we do not assume scalar alignment,
which means that, in the Higgs basis \eqref{expansion-Higgs-basis}, 
the physical Higgs boson $\hsm$ does not coincide with $\rho_1$
but is a mixture of all five neutral scalars. 
We parametrize it using four mixing angles $\epsilon$, $\alpha$, $\gamma_1$, and $\gamma_2$ in the following way:
\begin{equation}
	\hsm = \rho_1 c_\epsilon + s_\epsilon c_\alpha (c_{\gamma_1} \rho_2 + s_{\gamma_1} \rho_3) +  
	s_\epsilon s_\alpha (c_{\gamma_2} \eta_3 + s_{\gamma_2} \eta_2)\,.
	\label{hsm}
\end{equation}
The most important parameter is $\cos\epsilon$. It plays the same role as the famous $\sin(\beta-\alpha)$ in the 2HDM:
$h_{SM}VV$ couplings are proportional to $\cos\epsilon$, which is expected to be close to 1.
The value of $\alpha$ characterizes the admixture of the imaginary components of the neutral complex fields $\Phi^0_2$ vs. $\Phi^0_3$,
while the angles $\gamma_1$ and $\gamma_2$ parametrize their relative weights.
These angles directly enter the expressions for the quark couplings with the SM-like Higgs
and, therefore, can be probed experimentally.

With these conventions, we use the following 12 quantities as input:
\begin{eqnarray}
	\mbox{vev alignment:} && v\,, \quad \beta\,, \quad \psi\,, \nonumber\\
	\mbox{SM-like Higgs properties:} && \msm\,, \quad \epsilon\,, \quad \alpha\,, \quad \gamma_1\,, \quad \gamma_2\,, \nonumber\\
	\mbox{extra parameters:} && m_{H_1^\pm}^2\,, \quad m_{H_2^\pm}^2\,, \quad m_{11}^2 - m_{22}^2\,,\quad \lambda_{89}\,.  \label{input}
\end{eqnarray}
The last two parameters, although not directly observables, play important roles. 
It is clear from Eqs.~\eqref{M-Higgs-basis} and \eqref{aa-11-12}, setting $m_{11}^2 - m_{22}^2 = 0$ will automatically
lead to scalar alignment $\epsilon = 0$. This is a special case, for which the present inversion procedure
is not applicable, and a new version of the procedure must be constructed, which is discussed in Section~\ref{subsection-alignment}.
Although in the exact scalar alignment corresponds to $m_{11}^2 - m_{22}^2 = 0$ and $\epsilon = 0$,
it does not imply that numerically small $(m_{11}^2 - m_{22}^2)/v^2$ corresponds to equally small $\beta$.
The last parameter $\lambda_{89} \ge 0$ defined in Eq.~\eqref{lam89} quantifies mixing 
between the real and imaginary components.
Setting $\lambda_{89} = 0$ will lead to accidental symmetries, and the model is no longer CP4 but is $D_4$ 3HDM.
Thus, $\lambda_{89}$ encodes the proximity to this very special limit.

The main idea behind our algorithm of reconstructing the coefficients of the potential 
from the input quantities \eqref{input} is to rely on the very special --- tridiagonal --- form
of the neutral scalar mass matrix $\tilde {M}$ written in Eq.~\eqref{M-Higgs-basis}.
The SM-like Higgs $\hsm$, with its mass squared $\msm$ is an eigenvector of this matrix:
\begin{equation}
	\left(\begin{array}{ccccc}
		a_{11} & a_{12} & 0 & 0 & 0  \\
		a_{12} & a_{22} & a_{23} & 0 & 0  \\
		0 & a_{23} & a_{33} & a_{34} & 0  \\
		0 & 0 & a_{34} & a_{44} & a_{23}  \\
		0 & 0 & 0 & a_{23} & a_{55} 
	\end{array}\right)
	\left(\begin{array}{l}
		c_\epsilon \\
		s_\epsilon c_\alpha c_{\gamma_1}\\
		s_\epsilon c_\alpha s_{\gamma_1}\\
		s_\epsilon s_\alpha c_{\gamma_2}\\
		s_\epsilon s_\alpha s_{\gamma_2}
	\end{array}\right)
	= 	\msm \left(\begin{array}{l}
		c_\epsilon \\
		s_\epsilon c_\alpha c_{\gamma_1}\\
		s_\epsilon c_\alpha s_{\gamma_1}\\
		s_\epsilon s_\alpha c_{\gamma_2}\\
		s_\epsilon s_\alpha s_{\gamma_2}
	\end{array}\right)\,.\label{SM-vector}
\end{equation}
By explicitly writing the five equations, we can algorithmically express the entries of this mass matrix
in terms of the input angles. Since we have eight independent $a_{ij}$ and only five relations,
it seems at first that we just need to specify three out of eight $a_{ij}$ as additional input parameters.
However, $a_{ij}$ themselves are non-trivial, correlated functions of the input parameters Eq.~\eqref{input},
thus cannot be chosen arbitrarily. 

It is here that the tridiagonal form of the mass matrix helps overcome the difficulty.
Since the first and the last rows involve only two entries each, one is able to reconstruct 
the coefficients of the potential
through a {\em linear chain of relations} using the input parameters in Eq.~\eqref{input}. 
In Appendix~\ref{appendix-algorithm}, we provide the details of the procedure and the explicit expressions 
for all the parameters of the potential.
The last step is to insert all these quantities in the charged Higgs mass matrix Eq.~\eqref{mass-matrix-charged-Higgs}
and use the two charged Higgs masses $m_{H_1^\pm}^2$ and $m_{H_2^\pm}^2$ to determine
the remaining free parameters $\lambda_4$ and $\lambda_4'$.

With this algorithm, we build the scalar sector which, by construction, 
has the desired vevs and all the properties of the SM-like Higgs $\hsm$.
We stress that this procedure is linear and, for any choice of the input parameters,
it produces numerical values of the coefficients of the potential.
Since no freedom is left in the potential, the properties of all four additional 
neutral scalars, including their masses, follow uniquely and cannot be adjusted. 
To recover them, we consider again the neutral Higgs mass matrix \eqref{M-Higgs-basis} 
and diagonalize it numerically, finding its eigenvectors and eigenvalues.
As a cross check we verify that we recover the SM-like Higgs 
and explore the spectrum of the other four neutrals.

	\subsection{The origin of the tridiagonal mass matrix}

To get an insight into the origin of the all-important tridiagonal structure 
of the neutral mass matrix, Eq.~\eqref{M-Higgs-basis},
we rewrote the starting Higgs potential in the Higgs basis, in terms of $\Phi_i$ defined in 
Eq.~\eqref{Phiphi}, and took into account the relations Eq.~\eqref{lam89}--\eqref{t23}.
The full potential takes in this basis a rather cumbersome form.
However, we can focus on the terms which eventually give rise to the neutral mass matrix.
They include the quadratic terms of the potential,
\begin{eqnarray}
	V_2 &=& - m_{22}^2\left(\Phi_1^\dagger\Phi_1 + \Phi_2^\dagger\Phi_2 +\Phi_3^\dagger\Phi_3\right)\nonumber\\
	&&
	- (m_{11}^2 - m_{22}^2)\left[c_\beta^2 \Phi_1^\dagger\Phi_1 + s_\beta^2 \Phi_2^\dagger\Phi_2 
	- 2c_\beta s_\beta\Re(\Phi_1^\dagger\Phi_2) \right]\,,
	\label{quadratic-Higgs}
\end{eqnarray}
as well as the quartic terms with at least two $\Phi_1$'s or its conjugates.
From these expressions, several observations could be made.

First, we verified that all terms of the type $(\Phi_1^\dagger\Phi_1)(\Phi_1^\dagger\Phi_3)$ cancel
due to the relation \eqref{t23}. This is not surprising. Since $\Phi_1^0$ contains the vev, this term, if present, 
would leads to the tadpole terms for the field $\Phi_3^0 = (\rho_3 + i \eta_3)/\sqrt{2}$,
which are not compensated by $\Phi_1^\dagger\Phi_3$ in the quadratic part \eqref{quadratic-Higgs}.
As a result, we conclude that $\rho_1$ does not mix with $\rho_3$ or $\eta_3$.
In fact, $\rho_1$ can only mix with $\rho_2$ because only $\Re(\Phi_1^\dagger\Phi_2)$
is present in the quadratic part.
In a similar way, we find the terms $(\Phi_1^\dagger\Phi_1)\,\Re(\Phi_2^\dagger\Phi_3)$
but not $(\Phi_1^\dagger\Phi_1)\,\Im(\Phi_2^\dagger\Phi_3)$,
which implies that $\rho_2$ mixes with $\rho_3$ and $\eta_2$ mixes with $\eta_3$.
Finally, we also observe terms of the type $\lambda_{89}\Re(\Phi_1^\dagger\Phi_3) \Im(\Phi_1^\dagger\Phi_3)$,
which allow for $CP$-violating mixing between $\rho_3$ and $\eta_3$.

What emerges from these observations is a very particular mixing patterns in the mass terms:
$\Phi_1 \leftrightarrow \Phi_2 \leftrightarrow \Phi_3$, with $\Phi_3$ allowing for $(\rho_3,\eta_3)$ mixing.
It is precisely this pattern that is encoded in the tridiagonal form of the neutral mass matrix:
$\rho_1 \leftrightarrow \rho_2 \leftrightarrow \rho_3\leftrightarrow \eta_3\leftrightarrow \eta_2$.

It is not immediately clear whether this feature is a peculiarity of the CP4 3HDM
or can be connected with a general structural feature of scalar potentials.
For a model to exhibit the tridiagonal neutral mass matrix, 
we must make sure, at the very least, that the quadratic terms are very limited 
and include, in the Higgs basis, only one type of interactions between $\Phi_1$ and another doublet.
However, this is not sufficient, as the quartic terms need to satisfy additional conditions.
Nevertheless, we believe that other multi-Higgs models exist that follow the same pattern.

\section{The scalar sector of the CP4 3HDM: numerical exploration}\label{section-numerical}

In this section, we present numerical results and qualitative features which
we observed when implementing the above algorithm.
However, instead of presenting just the final results,
we would like to explain how our understanding of the scalar sector of the CP4 3HDM evolved.
We will first present a general scan in a wide range of the input parameters
and describe the main features which emerged from the plots.
However, as we will see, a similar scan in the alignment limit seems to give a very different picture
than the general scan close to the alignment limit.
This surprising mismatch prompted us to take a deeper look into the properties of the mass matrix
and helped us understand why, in our initial general scan, 
we missed an all-important region in the parameter space in which all extra Higgses 
are significantly heavier than the $\hsm$.
We then correct this omission by performing a focused scan and explore the results.

\subsection{Procedure and constraints}

The algorithm outlined above and detailed in Section~\ref{appendix-algorithm} 
was implemented in three independent Python codes, which agreed on the results.
We performed different kinds of scans, fixing the values of a subset of parameters
and scanning over the others.

When implementing the algorithm, we imposed the following constraints:
\begin{itemize}
	\item
	Starting from the input parameters \eqref{input}, we obtain the set of $\lambda_i$.
	Since the reconstruction algorithm is linear, a solution always exists and is unique.
	\item 
	Knowing all $\lambda_i$, we build the $5\times 5$ neutral mass matrix $\tilde{\cal M}$, 
	numerically diagonalize it, and check that its eigenvalues are all positive. 
	If we encounter at least one negative eigenvalue, we drop this solution, as it corresponds to a saddle point, 
	and choose another set of input parameters.
	Since numeric diagonalization is computer-time consuming, we first perform a quick check of the positivity 
	of the diagonal elements $\tilde{\cal M}$. 
	It turns out that, for most saddle points, this quick check identifies the problem,
	which allows us to skip numerical diagonalization and drop this point.
	\item
	Once all the neutral Higgs masses are known, we choose a set of charged Higgs boson masses 
	and make sure that the electroweak precision parameters $S$, $T$, and $U$  
	are within the $2\sigma$ range of their experimental values. 
	For this, we used the expressions from \cite{Grimus:2008nb}. 
	\item
	Having the set of $\lambda_i$, we check whether they satisfy the bounded-from-below (BFB) conditions 
	and the unitarity and perturbativity constraints.
	Since the exact necessary and sufficient BFB conditions are not known for the CP4 3HDM,
	we used the sufficient conditions described in \cite{Ferreira:2017tvy}.
	For the unitarity bounds, we implemented the algorithm presented in \cite{Bento:2022vsb}.
\end{itemize}

The main quantities of interest are the masses of the extra Higgs bosons,
in particular, the minimal $\mmin$ and the maximal $\mmax$ masses 
among the four additional neutral Higgs bosons.
Since CP4 3HDM does not possess the decoupling limit, we need to understand 
how far along the mass scale we can in principle push the extra Higgses
without conflicting the constraints and to which regions
in the input parameter space such high masses point.

\subsection{A general scan}

Let us now illustrate the main results of a generic scan in the parameter space within the following ranges:
\begin{eqnarray}
&&\tan\beta \in [0.1, 10]\,, \quad \tan\psi \in [0.1, 10]\,, \quad 
\cos\epsilon > 0.8\,, \quad \alpha, \gamma_1, \gamma_2 \in [0, 2\pi]\,, \label{generic-scan-range}\\
&& 
m_{11}^2 - m_{22}^2 \in [-1, 1]\times (300\,\GeV)^2\,, \quad m_{H_{1,2}^\pm}^2 \in [0, (700\,\mbox{GeV})^2]\,,
\quad \lambda_{89} \in (0, 4)\,.
\nonumber
\end{eqnarray}
The values of $v=246$~GeV and $m_{\mbox{\tiny SM}} = 125$~GeV are fixed.

\begin{figure}[h]
	\centering
	\includegraphics[height=7.5cm]{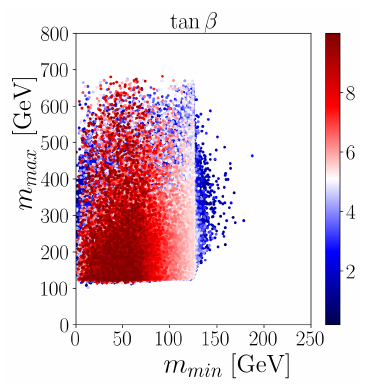}\hfill
	\includegraphics[height=7.5cm]{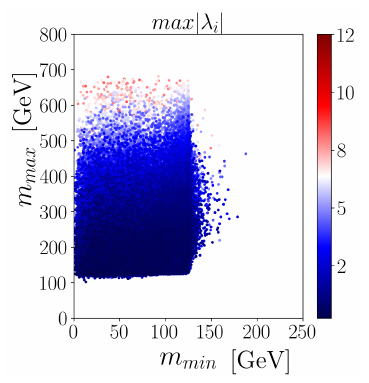}
	\caption{The results of a general scan within the region given in Eq.~\eqref{generic-scan-range} 
		projected on the $(\mmin,\mmax)$ plane.
		Color indicates $\tan\beta$ (left plot) and the maximal value of $|\lambda_i|$ (right plot).}
	\label{fig-scan-1}
\end{figure}
In Fig.~\ref{fig-scan-1}, we show the results of this scan with $N = 10^5$ random points
projected on the plane $(\mmin,\mmax)$.
In the left pane, color encodes the values of $\tan\beta$, with the large $\tan\beta$ points shown on top.
We observe several characteristic features.
The entire region covered by points contains two characteristic subregions.
The large $\tan\beta$ points, shown in shades of red, cover an approximately rectangular shape with $\mmin \lsim m_{\mbox{\tiny SM}}$ and
$100\,\GeV \lsim \mmax \lsim 650\,\GeV$. The larger $\tan\beta$, the lower is the mass $\mmin$ 
of the lightest neutral Higgs boson. 

On the other hand, when $\tan\beta \sim 1$, an additional region opens up in which
all the neutral scalars are heavier than the SM-like Higgs.
Still, the plot seems to indicate that there exists a upper limit on the lightest additional Higgs boson,
which is rather low, around 200~GeV. This low value together with the potentially unsuppressed
FCNC in the quark sector \cite{Zhao:2023hws} can lead one to suspect that all these point
will eventually run into conflict with experimental data, once the full phenomenological scan is performed.

Fig.~\ref{fig-scan-1} also displays an upper bound of about 650~GeV on all the additional neutral Higgses.
This upper bound is a manifestation of the absence of the decoupling limit in this model.
Namely, it is impossible to render the extra Higgses arbitrarily heavy by adjusting $m_{ii}^2$ alone;
to achieve that, one would need to increase some of $|\lambda_i|$ to unacceptably large values.
To demonstrate this dependence, we show in the right pane of Fig.~\ref{fig-scan-1} the same distribution
but with color indicating the maximal value among all $|\lambda_i|$ from the potential.
In this plot, points with smaller values of $|\lambda_i|_{\rm \max}$ are shown on top of larger
values of $|\lambda_i|_{\rm \max}$. As we see, pushing $\mmax$ above 600~GeV 
requires some $|\lambda_i|$ to grow to 10 and beyond.

\begin{figure}[h]
	\centering
	\includegraphics[height=7.5cm]{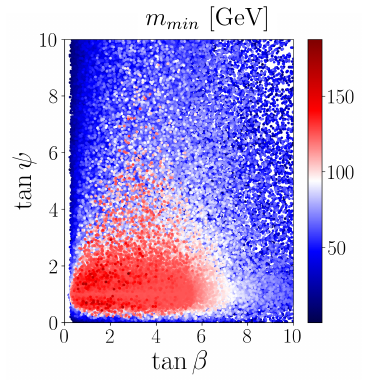}
	\caption{The results of a general scan projected on the $(\tan\beta, \tan\psi)$ plane, with color indicating $\mmin$.}
	\label{fig-scan-2}
\end{figure}

Fig.~\ref{fig-scan-2} presents another insight of the general scan results. 
Here, the points are now projected on the $(\tan\beta, \tan\psi)$ plane,
with color indicating the value of $\mmin$. 
Although the points cover the entire range scanned,
most of them leads to dangerously light extra Higgs bosons.
In particular, if we are interested in the situation where the SM-like Higgs is the lightest neutral scalar,
this plot clearly favors the region of moderate $\tan\beta$ and $\tan\psi$.

These plots together with other projections which we have studied give us several lessons.
A random choice of the input parameters typically leads to a light neutral Higgs boson, 
with mass below 100~GeV. 
In addition, quite often it generates one or several additional neutral Higgses lying close 
to 125~GeV. As we checked, more than half of the general scan points contain
an extra Higgs within the range 120 to 130~GeV.

Certainly, phenomenology of these light scalars strongly depends on the values of the mixing angles,
in particular on $\cos\epsilon$, as well as on their couplings with fermions. 
However, by default, such light Higgses are considered dangerous as they can easily run into conflict
with the LHC searches of new scalars. 
This is why we are interested in the cases when all the Higgses are sufficiently heavy.
But the general random scan with $10^5$ points seems to indicate that there always exist
a Higgs lighter than about 250~GeV.

\subsection{The puzzling results in the alignment limit}\label{subsection-alignment}

Next, we compare our results of a general random scan with the previous phenomenological exploration 
of the CP4 3HDM scalar sector reported in \cite{Ferreira:2017tvy}.
In that paper, two extra assumptions were used: exact scalar alignment and the assumption
that the observed 125~GeV Higgs is the lightest neutral scalar among the five physical Higgs bosons.

Scalar alignment corresponds to the situation when the SM-like Higgs $\hsm$ coincides,
in the Higgs basis, with $\rho_1$ in Eq.~\eqref{expansion-Higgs-basis}.
In terms of our input parameters, it is achieved by setting $\epsilon = 0$.
As it was already mentioned in \cite{Ferreira:2017tvy}, the only way to achieve this limit is to
set $m_{11}^2 - m_{22}^2 = 0$, which leads to $m_{11}^2 = m_{22}^2 = \msm$.

Since our algorithm does not apply to the case $\epsilon = 0$, we developed a different version 
of the parametrization designed specifically for the alignment limit.
As the SM-like Higgs decouples from the extra four,
we now use the mass of one of extra Higgses as a new free parameter $m_H^2$. 
Thus, the total count of the input parameters is eleven.
The mixing angles $\alpha$, $\gamma_1$, and $\gamma_2$ are defined similarly to the general scan
but now they corresponds not to the SM-like Higgs but to the new Higgs with the mass $m_H$:
\begin{equation}
	\left(\begin{array}{cccc}
		a_{22} & a_{23} & 0 & 0  \\
		a_{23} & a_{33} & a_{34} & 0  \\
		0 & a_{34} & a_{44} & a_{45}  \\
		0 & 0 & a_{45} & a_{55} 
	\end{array}\right)
	\left(\begin{array}{l}
		c_\alpha c_{\gamma_1}\\
		c_\alpha s_{\gamma_1}\\
		s_\alpha c_{\gamma_2}\\
		s_\alpha s_{\gamma_2}
	\end{array}\right)
	= 	m_H^2 \left(\begin{array}{l}
		c_\alpha c_{\gamma_1}\\
		c_\alpha s_{\gamma_1}\\
		s_\alpha c_{\gamma_2}\\
		s_\alpha s_{\gamma_2}
	\end{array}\right)\,.\label{H-aligned-vector}
\end{equation}	
The expressions for the $a_{ij}$ in terms of the parameters of the potential as the same as before.
The algorithm constructed above and described in detail in Appendix~\ref{appendix-algorithm}
can be repeated here, with the replacement $\msm \to m_H^2$. 

\begin{figure}[h]
	\centering
	\includegraphics[height=7.5cm]{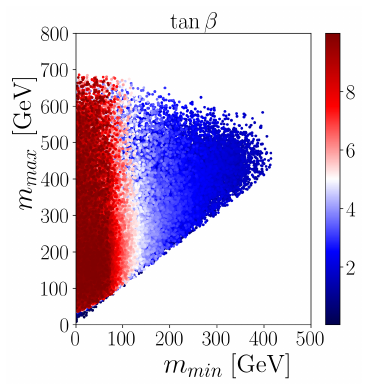}
	\caption{The results of a scan for the case of exact scalar alignment, with color indicating $\tan\beta$.}
	\label{fig-scan-aligned}
\end{figure}

We then perform a random scan in the parameter space of this alignment-based version of the algorithm,
where the new Higgs boson mass $m_H$ is scanned in the range from 0 to 450~GeV.
The results are presented in Fig.~\ref{fig-scan-aligned} 
on the $(\mmin,\mmax)$ plane, with color indicating the value of $\tan\beta$.

This plots differs in several aspects from the outcome of the general misaligned case of Fig.~\ref{fig-scan-1}, left.
The values of $\mmin$ now extend to about 420~GeV, which is in stark contrast with the approximate upper bound
of 200~GeV we previously found in a general scan.
In addition, there is now a small-mass corner, with all Higgses lighter than 100~GeV, which was absent in
our general scan.

This mismatch is surprising. It is natural to expect that a general scan, which extends up to $\cos\epsilon = 0.9999$,
should smoothly pass to the alignment case, $\cos\epsilon = 1$, 
so that our previous plot Fig.~\ref{fig-scan-1}, left, should cover the entirety of the new plot Fig.~\ref{fig-scan-aligned}.
However, our plots show that this is clearly not the case.
We conclude that our general scan, despite testing $10^5$ random points, 
missed two key regimes of this model.
In particular, the general scan missed the all-important and phenomenologically promising 
high-mass region, in which all the neutral Higgses are in the range of 300--600~GeV.

The only explanation to this puzzling observation is that these high-mass and low-mass regions,
when mapped to the space of the observable-driven input parameters \eqref{input},
have extremely small measure. Within our linear inversion algorithm, 
they correspond to certain cancellations among various terms which eventually drive all the extra Higgs
masses either to high or to low values.
As a result, the random scan points
uniformly distributed within the ranges \eqref{generic-scan-range} simply miss these regions.
However, with a different parameter choice used in the alignment limit scan,
these regions are no longer tiny, as illustrated by Fig.~\ref{fig-scan-aligned}.

\subsection{The resolution of the puzzle}

This brings us back to the question of a suitable strategy for choosing the input parameters
\eqref{input} which would be able to reveal the all-important high-mass region missing in Fig.~\ref{fig-scan-1}. 

To address this question, let us first discuss a useful property of tridiagonal matrices.
Any symmetric tridiagonal matrix can be decomposed into a sum of $2\times 2$ block diagonal matrices of the following type:
\begin{eqnarray}
&&	\left(\begin{array}{ccccc}
		a_{11} & a_{12} & 0 & 0 & \cdots  \\
		a_{12} & a_{22} & a_{23} & 0 & \cdots \\
		0 & a_{23} & a_{33} & a_{34} & \cdots \\
		0 & 0 & a_{34} & a_{44} & \cdots \\
		\cdots & \cdots & \cdots & \cdots & \cdots 
	\end{array}\right)
=  a_0 \cdot \id + a_{12} 
	\left(\begin{array}{ccccc}
	\tan x_2 & 1 & 0 & 0 & \cdots  \\
	1 & \cot x_2 & 0 & 0 & \cdots \\
	0 & 0 & 0 & 0 & \cdots \\
	0 & 0 & 0 & 0 & \cdots \\
	\cdots & \cdots & \cdots & \cdots & \cdots 
\end{array}\right)\label{tridiag-decomposition}\\
&& \qquad \qquad
+\ a_{23} 
\left(\begin{array}{ccccc}
	0 & 0 & 0 & 0 & \cdots  \\
	0 & \tan x_3 & 1 & 0 & \cdots \\
	0 & 1 & \cot x_3 & 0 & \cdots \\
	0 & 0 & 0 & 0 & \cdots \\
	\cdots & \cdots & \cdots & \cdots & \cdots 
\end{array}\right) 
+\ a_{34} 
\left(\begin{array}{ccccc}
	0 & 0 & 0 & 0 & \cdots  \\
	0 & 0 & 0 & 0 & \cdots \\
	0 & 0 & \tan x_4 & 1 & \cdots \\
	0 & 0 & 1 & \cot x_4 & \cdots \\
	\cdots & \cdots & \cdots & \cdots & \cdots 
\end{array}\right) + \dots\nonumber
\end{eqnarray}
With this parametrization, we effectively trade the diagonal elements $a_{ii}$, $i > 1$,
for the ``angles'' $x_i$.
Although finding the eigenvalues of the matrix is still a difficult task,
one can easily identify the regime in which all the eigenvalues apart from $a_0$ 
become parametrically large even for moderate $a_{ij}$:
one simply needs to choose all $\tan x_i \ll 1$.

To make use of this observation, let us rewrite the $5\times 5$ neutral mass matrix $\tilde {\cal M}$
via the input parameters:
\begin{eqnarray}
	\tilde {\cal M}& = & \msm \cdot \id_5 \ + \ \sin2\beta \,(m_{11}^2-m_{22}^2)
	\left(\begin{array}{ccccc}
		c_\alpha c_{\gamma_1} \tan\epsilon & -1 & 0 & 0 & 0  \\
		-1 & (c_\alpha c_{\gamma_1} \tan\epsilon)^{-1} & 0 & 0 & 0 \\
		0 & 0 & 0 & 0 & 0 \\
		0 & 0 & 0 & 0 & 0 \\
		0 & 0 & \ 0\  & \ 0\  & \ 0\  
	\end{array}\right)\nonumber\\[2mm]
	&-&\frac{\msm \, c_\beta s_{\gamma_2}}{c_\beta c_{\gamma_2} + s_{\gamma_2}t_{2\psi}} 
	\left(\begin{array}{ccccc}
		0 & 0 & 0 & 0 & 0 \\
		0 & \tan \gamma_1 & -1 & 0 & 0 \\
		0 & -1 & \cot \gamma_1 & 0 & 0 \\
		0 & 0 & 0 & \tan\gamma_2 & -1 \\
		0 & 0 & 0 & -1 & \cot\gamma_2
	\end{array}\right) \nonumber\\[2mm]
	&+ &\lambda_{89} v^2 s_\beta^2  
	\left(\begin{array}{ccccc}
		0 & 0 & 0 & 0 & 0 \\
		0 & 0 & 0 & 0 & 0 \\
		0 & 0 & (c_{\gamma_2}/s_{\gamma_1})\tan \alpha & -1 & 0 \\
		0 & 0 & -1 & (s_{\gamma_1}/c_{\gamma_2})\cot \alpha & 0 \\
		\ 0\  & \ 0\  & \ 0\  & \ 0\  & \ 0\  
	\end{array}\right) \,.\label{MH-decomposition}
\end{eqnarray}
This decomposition is of the form \eqref{tridiag-decomposition}.
Thus, in order to drive the masses of the extra Higgses,
we should focus on the region
\begin{equation}
	|\tan\epsilon| \ll 1\,, \quad |\tan\alpha| \ll |\tan\gamma_1| \ll 1\,, \quad |\tan\gamma_2| \ll 1\,.
	\label{small-angles}
\end{equation}
At the same time, the coefficients in front of the matrices should not be too small.
This mean, in particular, that $(m_{11}^2-m_{22}^2)/v^2$, $\lambda_{89}$, 
and $\tan\beta$ should all be of order 1.
The coefficient in the second line of Eq.~\eqref{MH-decomposition}
must also be positive and not too small. At first, it may seem problematic due to the $s_{\gamma_2}$ factor.
However, it can be compensated by the denominator which can be made small by adjusting the angle $\psi$.
In particular, it means that we should focus on the values $|\tan2\psi| \gg 1$ in order to overcome the small $s_{\gamma_2}$,
which gives preference for $\psi$ around $\pi/4$ but not exactly equal to this value.

When adjusting the mixing angles as in Eq.~\eqref{small-angles}, we must always control the magnitude of $\lambda_i$
emerging from the inversion algorithm. Tiny mixing angles can drive masses to high values but, at the same time,
render some $\lambda_i$ unacceptably large. We encounter once again the clash between
perturbativity and unitarity constraints with the desire to keep masses high.

We conclude that there indeed exists a way to make all extra Higgses significantly 
heavier than the SM-like scalar $\hsm$.
Although this region seems fine-tuned in the space of the input parameters,
this fine-tuning does not bear any physical significance
and, for a different choice of the free parameters, this region blows up.
However, we still prefer working with our input parameters because the inversion procedure is linear;
the benefits of linearity are higher than the price of scanning within a fine-tuned region of the parameter space.

\subsection{Focused scan in the high-mass region}

With the above insights, we can now repeat the scan in the parameter space optimizing
the scanning range to access the high-mass region which we missed before.
In the new scan, we limit the most important input parameters to lie within the following limited ranges:
\begin{eqnarray}
	&& \tan\beta \in [0.5, 2]\,, \quad \tan\psi \in [0.5, 3]\,, \quad
|\tan\alpha| < 0.05\,,
	\quad |\tan\gamma_{1}|, \ |\tan\gamma_{2}| < 0.3\,.\quad
	\label{focused-scan-range}
\end{eqnarray}
The other parameters are scanned within the same range as in Eq.~\eqref{generic-scan-range}.

\begin{figure}[h]
	\centering
	\includegraphics[height=7.5cm]{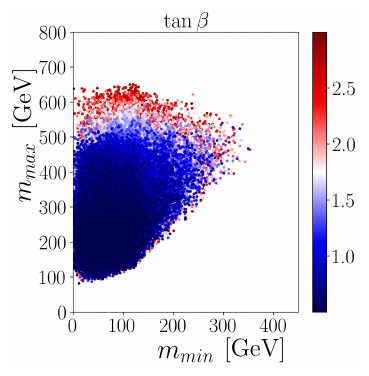}\hfill
	\includegraphics[height=7.5cm]{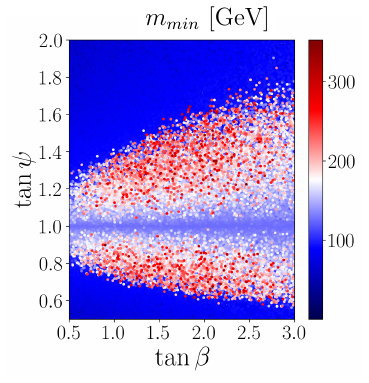}
	\caption{The results of the focused scan within the region given in Eq.~\eqref{focused-scan-range} 
		projected on the plane $(\mmin,\mmax)$ (left pane) and the plane $(\tan\beta, \tan\psi)$ (right pane).}
	\label{fig-scan-focused}
\end{figure}
In Fig.~\ref{fig-scan-focused}, we present the results of this focused scan.
In contrast with the general scan shown in Fig.~\ref{fig-scan-1}, the minimal Higgs mass can now extend
up to 350~GeV. In this figure, we plot the lower $\tan\beta$ points on top of higher $\tan\beta$
because the very high $\tan\beta$ range, which appeared as dark red in Fig.~\ref{fig-scan-1}, is now cut off.
The right plot in Fig.~\ref{fig-scan-focused}, to be compared with Fig.~\ref{fig-scan-2}, makes it clear such high values of $\mmin$ 
indeed appear in the region of $\tan\psi$ in the vicinity of 1, but not too close to 1,
just as described after Eq.~\eqref{small-angles}. 

To illustrate these results, we provide a benchmark point with sufficiently high masses.
The input parameters are
\begin{eqnarray}
&& \tan\beta = 0.8980, \quad \tan\psi = 0.8987,\quad \cos\epsilon = 0.8664,\nonumber\\
&&\alpha = -0.01023,\quad \gamma_1 = -0.05693,\quad \gamma_2 = -0.09003, \quad \lambda_{89} = 0.9640, \nonumber\\
&&m_{11}^2 - m_{22}^2 = 46652.7\,\GeV^2,\quad 
m_{H_1^\pm} = 288.6\,\GeV\,, \quad m_{H_2^\pm} = 337.5\,\GeV\,,
\label{benchmark-a}
\end{eqnarray}
which results in the following values of the potential:
\begin{eqnarray}
&&	m_{11}^2 = 42013.2\,\GeV^2, \quad m_{22}^2 = -4639.5\,\GeV^2, \quad 
\lambda_1 = 1.685,\quad \lambda_2 = 2.863,\nonumber\\
&&\lambda_{34} = 1.118,\quad \lambda'_{34} = 2.333,\quad \lambda_5 = -2.194,\quad 
\Re\lambda_8 = -2.939,\quad \Re\lambda_9 = -0.797,\nonumber\\
&& 
\lambda_3 = 1.719, \quad \lambda_3' = 1.129,\quad \lambda_4 = -0.602,\quad \lambda_4' = 1.203\,.	
\end{eqnarray}
This potential leads to the following masses of the four additional neutral scalars $H_2$ through $H_5$: 
\begin{equation}
	m_{H_i} \quad = \quad 349.2\,\GeV,\quad 357.7\,\GeV, \quad 392.8\,\GeV,\quad 467.0\,\GeV\,.
\end{equation}

\section{Discussion and conclusions}

In this work, we explored the scalar sector of the CP4 3HDM, a unique three-Higgs-doublet model with
a CP symmetry of order 4, which is spontaneously broken upon minimization of the potential. 
This model does not possess the decoupling limit, so it is important to find a regime
in which the additional scalars apart from the SM-like Higgs are sufficiently heavy.
Looking for this regime, we observed novel features of the model which were not noticed 
in the previous phenomenological study \cite{Ferreira:2017tvy} and which significantly facilitate the numerical analysis.

First, we found that the $5\times 5$ mass matrix for the neutral scalars, when rotated to the Higgs basis,
has a remarkable tridiagonal form, with non-zero elements only on the main diagonal 
and on the parallel diagonals just above and below it. 
The tridiagonal matrix admits a special decomposition 
which allows us to identify which parameters need to be large or small in order to obtain
all extra neutral Higgses significantly heavier than the SM-like Higgs.

Second, we constructed an improved scanning procedure, in which several physical quantities,
rather than the coefficients of the potential, are used as input.
Such a procedure, which we call ``inversion'', is routinely used in the 2HDM, but it is in general impractical
beyond two doublets. We showed that the CP4 3HDM admits a {\em linear} inversion, which dramatically 
speeds up the random scan of the model.
The feature that allows for linear inversion is again the tridiagonal form of the mass matrix.

These features based on the tridiagonal mass matrix are universal and may be helpful
in exploration of other BSM models.

Third, we identified the promising region in the input parameter space in which
all extra Higgses lie in the mass range of 300--600~GeV. We believe that this regime can contain
benchmark models which would pass all the LHC and flavor physics constraints.
Moreover, due to curious correlations between the scalar and Yukawa sectors, we expect to predict
novel signatures for collider searches.
These issues can now be addressed with a full phenomenological scan of the model,
including scalar and Yukawa sectors and taking into account the LHC and flavor physics constraints.
Results of this full scan will be reported elsewhere.

\section*{Acknowledgments}

We are thankful to Antonio Morais and Roman Pasechnik for insightful discussions and reading the manuscript.
The work of B.L. and I.P.I. was supported by Guangdong Natural Science Foundation (project No.~2024A1515012789).
J.G. is supported by the Center for Research and Development in Mathematics and Applications (CIDMA) through the Portuguese Foundation for Science and Technology (FCT - Funda\c{c}\~{a}o para a Ci\^{e}ncia e a Tecnologia), references UIDB/04106/2020 (\url{https://doi.org/10.54499/UIDB/04106/2020}) and UIDP/04106/2020 (\url{https://doi.org/10.54499/UIDP/04106/2020}). J.G. is also supported by the projects with references CERN/FIS-PAR/0019/2021 (\url{https://doi.org/10.54499/CERN/FIS-PAR/0019/2021}), CERN /FIS-PAR/0021/2021 (\url{https://doi.org/10.54499/CERN/FIS-PAR/0021/2021}), CERN/FIS-PAR/0019/2021 (\url{https://doi.org/10.54499/CERN/FIS-PAR/0019/2021}) and CERN/FIS-PAR/0025/2021 (\url{https://doi.org/10.54499/CERN/FIS-PAR/0025/2021}).
J.G. is also directly funded by FCT through the doctoral program grant with the reference 2021.04527.BD (\url{https://doi.org/10.54499/2021.04527.BD}).

\appendix

\section{Scalar mass matrices in the original basis}\label{appendix-mass-matrices}

Working in the original basis given by Eq.~\eqref{expansion}, we can write the charged scalar mass terms as 
$h_i^- ({\cal M}_{ch})_{ij} h_j^+$, where 
\begin{equation}
	{\cal M}_{ch} = {v^2\over 2}\mmmatrix{-s_\beta^2(\lambda_4 + \lambda_5 s_{2\psi})}%
	{c_\beta s_\beta (\lambda_4 c_\psi + \lambda_5 s_\psi)}{c_\beta s_\beta (\lambda_4 s_\psi + \lambda_5 c_\psi)}%
	{\cdot }{\tilde\Lambda s_\beta^2 s_\psi^2 -\lambda_4 c_\beta^2}%
	{- \tilde\Lambda s_\beta^2 s_\psi c_\psi - \lambda_5 c_\beta^2}%
	{\cdot}{\cdot}%
	{\tilde\Lambda s_\beta^2 c_\psi^2 - \lambda_4 c_\beta^2}\,,\label{mass-matrix-charged}
\end{equation}
with the dots indicating the duplicated entries of this symmetric matrix, where
$\tilde\Lambda$ was defined in Eq.~\eqref{tildeLam}.
One can explicitly verify that the charged Goldstone boson, which corresponds to the
combination of fields given by 
$G^\pm = c_\beta h_1^\pm + s_\beta c_\psi h_2^\pm + s_\beta s_\psi h_3^\pm$
is an eigenvector of this matrix with the zero eigenvalue, as expected.
For the neutral scalars, the $6\times 6$ mass matrix can be written via $3\times 3$ blocks:
\begin{equation}
	{\cal M}_n = \mmatrix{M_{h}}{M_{ha}}{M_{ha}^T}{M_{a}}\,,
	\label{neutralMM}
\end{equation}
where $M_{h}$ describes couplings within the $h_i$ sector,
$M_{a}$ corresponds to the $a_i$ sector, and $M_{ha}$ describes their mixing.
Since $CP$ is spontaneously broken, we expect the mixing to be driven by the 
imaginary parts of $\lambda_8$ and $\lambda_9$, that is, by the non-zero $\lambda_{89}$
defined in Eq.~\eqref{lam89}.
The explicit expressions for these blocks are:
\begin{eqnarray}
	\frac{M_h}{v^2} &=&
	\mmmatrix{2\lambda_1 c_\beta^2}{\lambda_{34}c_\beta s_\beta \, c_\psi }{\lambda_{34} c_\beta s_\beta \,s_\psi}
	{\cdot}%
	{2\lambda_2 s_\beta^2\, c_\psi^2}%
	{2\lambda_2 s_\beta^2\, s_\psi c_\psi}
	{\cdot}%
	{\cdot}%
	{2\lambda_2 s_\beta^2\, s_\psi^2} +
	\mmmatrix{0}{\lambda_5 c_\beta s_\beta\, s_\psi}{\lambda_5 c_\beta s_\beta\, c_\psi}
	{\cdot}%
	{\Lambda s_\beta^2\, s_\psi^2}%
	{-\Lambda s_\beta^2\, s_\psi c_\psi - \lambda_5 c_\beta^2}
	{\cdot}%
	{\cdot}%
	{\Lambda s_\beta^2\, c_\psi^2} \nonumber\\
	&&+ {1\over 2}\Re \lambda_9\, s_\beta^2\, t_{2\psi}
	\mmmatrix{0}{0}{0}{\cdot}{3c_{2\psi}-1}{3s_{2\psi}}{\cdot}{\cdot}{-3c_{2\psi}-1}\,,\label{mhh}\\
	\frac{M_a}{v^2} &=& \mmmatrix{-\lambda_5 s_\beta^2 s_{2\psi}}{\lambda_5 c_\beta s_\beta s_\psi}{\lambda_5 c_\beta s_\beta c_\psi}%
	{\cdot}{\Lambda' s_\beta^2 s_\psi^2}{- \lambda_5 c_\beta^2 - \Lambda' s_\beta^2 s_\psi c_\psi}%
	{\cdot}{\cdot}{\Lambda' s_\beta^2 c_\psi^2}\,,\label{maa}\\
	\frac{M_{ha}}{v^2} &=& - \lambda_{89} s_\beta^2 \mmmatrix{0}{0}{0}{0}{s_\psi^2}{-s_\psi c_\psi}{0}{-s_\psi c_\psi}{c_\psi^2}\,,
	\label{mha}
\end{eqnarray}
where $\Lambda'$ is defined in Eq.~\eqref{LamLam}.
The neutral Goldstone boson given by $G^0 = c_\beta a_1 + s_\beta c_\psi a_2 + s_\beta s_\psi a_3$
is also an eigenvector of the neutral mass matrix ${\cal M}_n$ with the zero eigenvalue.
These mass matrices coincides with the results of \cite{Ferreira:2017tvy}.

\section{The inversion algorithm: general case}\label{appendix-algorithm}

Working in the real vev basis, we use the following 12 parameters as input:
\begin{equation}
	v\,, \quad \beta\,, \quad \psi\,, \quad \msm\,,\qquad
	\epsilon\,, \quad \alpha\,, \quad \gamma_1\,, \quad \gamma_2\,, \qquad 
	m_{11}^2 - m_{22}^2\,,\quad \lambda_{89}\,.  \label{params-4}
\end{equation}
Since the SM-like Higgs with the mass squared $\msm$ is an eigenvector of the neutral mass matrix,
we obtain the coupled linear equations \eqref{SM-vector}, which we can 
explicitly write as five equations:
\begin{eqnarray}
	(1): &\quad& a_{12} s_\epsilon c_\alpha c_{\gamma_1} = (\msm - a_{11})\,c_\epsilon \nonumber\\ 
	(2): &\quad& a_{12} c_\epsilon +  a_{23} s_\epsilon c_\alpha s_{\gamma_1} = (\msm - a_{22})\,s_\epsilon c_\alpha c_{\gamma_1}\nonumber\\ 
	(3): &\quad& a_{23} c_\alpha c_{\gamma_1} +  a_{34} s_\alpha c_{\gamma_2} = (\msm - a_{33})\,c_\alpha s_{\gamma_1}\nonumber\\ 
	(4): &\quad& a_{34} c_\alpha s_{\gamma_1} +  a_{23} s_\alpha s_{\gamma_2} = (\msm - a_{44})\,s_\alpha c_{\gamma_2}\nonumber\\ 
	(5): &\quad& a_{23} c_{\gamma_2} = (\msm - a_{55})\,s_{\gamma_2}\,.\label{five-equations}
\end{eqnarray}
From these equations and the expressions of $a_{ij}$ in terms of the coefficients of the potential
given in Eqs.~\eqref{aa-11-12}--\eqref{aa-34}, we can determine all the parameters through the following algorithm:
\begin{enumerate}
	\item 
	Begin with the last equation in Eq.~\eqref{five-equations}, which directly relates $\gamma_2$ with $\lambda_5$:
	\begin{equation}
		\lambda_5 = - \frac{\msm}{v^2}\frac{s_{\gamma_2}}{c_\beta c_{2\psi} c_{\gamma_2} + s_{2\psi} s_{\gamma_2}}\,.
		\label{alg-lam5}
	\end{equation}
	Since we have already fixed $s_{2\psi} > 0$ and know that $\lambda_5 < 0$, 
	we find that the angles we select must satisfy:
	\begin{equation}
		c_\beta c_{2\psi} \cot\gamma_2 + s_{2\psi} > 0\,.\label{alg-condition1}
	\end{equation}
In principle, it implies that we should pick up the value of $\gamma_2$ from a certain interval which depends on $\beta$
and $\psi$. An alternative strategy, which we used in our code, is to pick up
a random $\gamma_2$ and simply check whether the inequality \eqref{alg-condition1} holds.
	
	\item
	The first equation in Eq.~\eqref{five-equations} can be written as
	\begin{eqnarray}
		(m_{22}^2 - m_{11}^2)\, s_{2\beta}\, c_\alpha c_{\gamma_1}\tan\epsilon 
		&=& \msm - (m_{11}^2 + m_{22}^2) - (m_{11}^2 - m_{22}^2)c_{2\beta}\,.
	\end{eqnarray}
	From here, we determine
	\begin{equation}
		m_{11}^2 + m_{22}^2 = \msm + (m_{11}^2 - m_{22}^2)(s_{2\beta}\, c_\alpha c_{\gamma_1}\tan\epsilon - c_{2\beta})\,.
		\label{alg-m11m22}
	\end{equation}
	This allows us to calculate $m_{11}^2$ and $m_{22}^2$ separately.
	
	\item
	The second equation in Eq.~\eqref{five-equations} can be used to obtain $\lambda_{34}$:
	\begin{equation}
		\lambda_{34} = -\lambda_5\left(s_{2\psi} + c_\beta c_{2\psi}\tan\gamma_1\right)  - 
		\frac{(\msm)^2 - 2\msm(m_{11}^2 + m_{22}^2) + 4m_{11}^2 m_{22}^2}{v^2\left[\msm - (m_{11}^2 + m_{22}^2) - (m_{11}^2 - m_{22}^2)c_{2\beta}\right]}\,.
	\end{equation}
	\item
	Having determined $\lambda_{34}$, we use the first tadpole equation \eqref{m11} to determine $\lambda_1$.
	\item
	The third line of Eq.~\eqref{five-equations} and the third tadpole condition Eq.~\eqref{t23} can be used 
	to determine $\Lambda$ and $\Re\lambda_9$. After some algebra, we obtain
	\begin{equation}
		\Lambda = \frac{\msm}{v^2}\frac{c_{4\psi}}{s_\beta^2} + \lambda_{89} c_{4\psi}\tan\alpha\frac{c_{\gamma_2}}{s_{\gamma_1}}
		- \lambda_5\left[\frac{c_\beta^2}{s_\beta^2}s_{2\psi}\left(2+c_{4\psi}\right) - \frac{c_\beta}{s_\beta^2}c_{2\psi}c_{4\psi}\cot\gamma_1\right]\,.
	\end{equation}
	Having found $\Lambda$, we use the tadpole equation to find $\Re\lambda_9$:
	\begin{equation}
		\Re\lambda_9 
		= - \frac{\lambda_5c_\beta^2 c_{2\psi} + \Lambda s_\beta^2 s_{2\psi} c_{2\psi}}{s_\beta^2 c_{4\psi}}\,.
		\label{extr-lam9-2}
	\end{equation}
	\item
	Next, we use the fourth line of Eq.~\eqref{five-equations} to fix $\Re\lambda_8$:
	\begin{equation}
		\Re\lambda_8 = \frac{\Lambda}{2} - \frac{\Re\lambda_9t_{2\psi}}{2}  - \frac{1}{2s_\beta^2}\left[\lambda_5\left(c_\beta c_{2\psi}\tan\gamma_2 - c_\beta^2 s_{2\psi}\right) 
		+ \lambda_{89} s_\beta^2 \cot\alpha\frac{s_{\gamma_1}}{c_{\gamma_2}} + \frac{\msm}{v^2}\right] \,.
	\end{equation}
	\item
	From the second tadpole equation, Eq.~\eqref{m22}, we compute $\lambda_2$. 
	\item
	Finally, from the definition of $\Lambda$ in Eq.~\eqref{Lambda}, we compute $\lambda_{34}'$.
\end{enumerate}
Together with $\lambda_4$ and $\lambda_4'$, which are calculated from the charged Higgs masses, 
we completely reconstruct all the coefficients in the potential.


\end{document}